# Tuning magnetic, lattice, and transport properties in CoNb$_3$S$_6$ via Fe doping

Deepu Kumar[1], Joydev Khatua[2], Mukesh Suthar[3], Rajesh Kumar Ulaganathan[4], Jeonghun Kang[5], Hengbo Cui[5], Sihun Seong[6], Seo Hyoung Chang[6], Kee Hoon Kim[5], Raman Sankar[3,*], Maeng-Je Seong[1,*], and Kwang-Yong Choi[2,*]

[1]*Department of Physics and Center for Berry Curvature-based New Phenomena (BeCaP) Chung-Ang University, Seoul 06974, Republic of Korea*

[2]*Department of Physics, Sungkyunkwan University, Suwon 16419, Republic of Korea*

[3]*Institute of Physics, Academia Sinica, Nankang, Taipei 11529, Taiwan*

[4]*Centre for Nanotechnology, Indian Institute of Technology Roorkee, 247667, India*

[5]*Department of Physics and Astronomy, CeNSCMR, Seoul National University, Seoul 08826, Republic of Korea*

[6]*Department of Physics, Chung-Ang University, Seoul 06974, Republic of Korea*

**Abstract**

We present a comprehensive investigation of the effects of Fe doping on the lattice dynamics, magnetic ordering, and magneto-transport properties of the intercalated van der Waals antiferromagnets Co$_{1-x}$Fe$_x$Nb$_3$S$_6$ ($x$ = 0.1 and 0.3). Temperature- and polarization-dependent Raman scattering measurements reveal a pronounced blue shift of the 180 cm$^{-1}$ phonon mode with increasing Fe concentration, indicating enhanced sensitivity of lattice vibrations to Fe-induced structural and mass effects. While the temperature evolution of the phonon modes is dominated by conventional anharmonic phonon softening, subtle anomalies observed near the Néel temperature for $x$ = 0.1 point to weak spin-phonon coupling. Electrical transport and magnetic susceptibility data show clear signatures of the antiferromagnetic phase transitions at $T_N$ ~ 20.5-23.7 K for $x$ = 0.1 and $T_N$ ~ 32.0 K for $x$ = 0.3. Out-of-plane magnetization measurements reveal hysteretic behavior with two field-induced transitions for $x$ =0.1, which evolve into a single hysteresis loop at $x$ =0.3, signaling a subtle reconstruction of the magnetic ground state. Magneto-transport measurements for $x$ = 0.1 further display a butterfly-shaped hysteretic magnetoresistance and a weak topological Hall effect; however, both features are strongly suppressed at $x$ = 0.3. These results illustrate the critical role of Fe-induced magnetic structure reconstruction in fine-tuning topological and magnetic transport phenomena in intercalated van der Waals antiferromagnets.

*Email: choisky99@skku.edu (K.Y. Choi), mseong@cau.ac.kr (M. J. Seong) and sankarndf@gmail.com (R. Sankar)

# I. INTRODUCTION

Two-dimensional (2D) van der Waals (vdW) transition-metal dichalcogenides (TMDCs) with the chemical formula $MX_2$ have attracted significant attention due to their rich and tunable physical properties. These include optical/excitonic phenomena [1], spin-valley coupling [2,3], superconductivity [4–6], magnetism [7], charge density waves [6], and many of which are strongly dependent on layer thickness and dimensionality [8,9]. Beyond dimensional control, intercalation of transition-metal atoms into the vdW gaps of layered TMDCs has proven to be an effective route to engineer novel electronic and magnetic states that are absent in pristine $MX_2$ compounds [10–16].

A particularly outstanding class of intercalated materials is described by the general formula $T_xMX_2$ (where T = V, Cr, Mn, Co, Ni; M = Ti, Nb, Ta; and X = S, Se, with $x$=1/3 or 1/4) [10,11,14–16]. Owing to strong interactions between the intercalated magnetic ions and the host lattice, these compounds exhibit physical properties that differ markedly from those of their parent $MX_2$ materials [4,17,18] and are highly sensitive to the intercalant concentration [10,19]. For instance, $Fe_{1/3}NbS_2$ exhibits long-range antiferromagnetic order with a Néel temperature $T_N \approx$ 42-45 K, whereas a slight reduction in Fe concentration to $Fe_{0.30}NbS_2$ drives the system into a spin-glass-like state [19]. In the literature, the fractional notation $T_xMX_2$ is also described using the $T_xM_3X_6$ nomenclature [18,20–23], which we primarily adopt throughout this work, unless otherwise specified.

Within this family, $CoNb_3S_6$, corresponding to $x = 1/3$ intercalation of Co atoms between $NbS_2$ layers, has attracted particular attention. $CoNb_3S_6$ is a collinear antiferromagnet with a Néel temperature of approximately $T_N \sim$ 26 K [24,25]. Notably, the parent compound 2H-$NbS_2$ is a superconductor below a critical temperature of about 6 K [26]; however, this superconducting state is completely suppressed upon intercalation with Co or other transition-metal atoms, highlighting the strong impact of interlayer magnetic ions on the electronic ground state. Remarkably, $CoNb_3S_6$ shows a pronounced anomalous Hall effect (AHE) within a narrow temperature window near the magnetic phase transition, under an external magnetic field applied along the *c*-axis [18]. This behavior has been attributed to the presence of weak ferromagnetic (FM) components and the interplay between nontrivial magnetic textures and the underlying topological electronic band structure.

Following the observation of a large AHE in $CoNb_3S_6$, *Mangelsen et al.* systematically investigated the relation between chemical composition and the AHE in $CoNb_3S_6$ in the magnetically ordered state, revealing that the AHE is strongly suppressed and nearly vanishes when the Co concentration exceeds the ideal stoichiometric value [10]. Subsequently, Su *et al.* reported the emergence of exchange bias and a topological Hall effect (THE) in $Co_{0.5}Fe_{0.5}Nb_3S_6$ [22]. Furthermore, *Zhu et al.* demonstrated that Fe doping enables manipulation of the magnetic structure and tuning of both the AHE and THE in $CoNb_3S_6$ [27]. These findings position $CoNb_3S_6$ as a compelling platform for exploring the coupling between magnetism and electronic topology in intercalated vdW materials.

Given the intriguing correlation among weak ferromagnetism, band topology, and stoichiometry, a systematic investigation of Fe doping over a broader and well-controlled concentration range remains highly desirable. Such a study would provide deeper insight into the microscopic mechanisms governing the emergence, evolution, and suppression of anomalous and topological transport phenomena in intercalated vdW antiferromagnets.

In this work, we report comprehensive Raman scattering, magnetic and electrical transport measurements of $Co_{1-x}Fe_xNb_3S_6$ with $x = 0.1$ and $0.3$. Our results demonstrate that Fe doping fine-tunes the lattice, magnetic, and electronic properties of $CoNb_3S_6$. As the Fe concentration increases toward $x = 0.3$, conventional antiferromagnetic behavior becomes dominant, while the topological Hall effect is progressively suppressed. These observations underscore the decisive role of a weak ferromagnetic structure in enabling topological transport behavior in intercalated vdW antiferromagnets.

**II. EXPERIMENTAL DETAILS**

Polycrystalline $Co_{1-x}Fe_xNb_3S_6$ ($x$= 0.1 and 0.3) was prepared by intimately mixing stoichiometric amounts of high-purity Co powder, Fe powder, Nb powder, and sulfur pieces. All sample preparation steps before quartz ampoule sealing were carried out in an argon-filled glove box, with oxygen and moisture levels maintained below < 0.5 ppm. The mixed powders were loaded into a quartz tube, evacuated, and flame-sealed before being placed in a tube furnace. The mixture was sealed in an evacuated quartz ampoule and heated at 920°C for 72 h to yield a homogeneous precursor. Approximately 5g of the resultant powder and 200 mg of iodine (I2) as the transport agent were loaded into a fused-silica tube (20 mm inner diameter, 40 cm length), which was then evacuated and flame-sealed. Single-crystal growth was carried out via chemical vapor transport in a horizontal two-zone tube furnace, with the source and growth zones maintained at 950°C and ca. 850°C, respectively, for 12 days. Plate-like single crystals with well-defined facets were obtained from the cold end of the ampoule.

Raman scattering measurements were performed in a backscattering geometry using a Princeton Instruments SpectraPro HRS-750 spectrometer with a 532 nm (2.33 eV) laser excitation. The incident laser power was limited to approximately 0.5 mW to minimize local heating effects. A 50× objective lens was used for both focusing the incident beam and collecting the scattered light. The scattered signal was dispersed using a 1200 grooves/mm grating and detected with an electrically cooled charge-coupled device (CCD). Temperature-dependent Raman spectra were acquired using a closed-cycle cryostat (Montana Instruments) over a temperature range of 3.5-330 K, with a temperature stability better than ± 0.1 K. At each temperature, a waiting time of approximately 10 min was allowed to ensure thermal equilibrium before data acquisition. Circular polarization-dependent Raman measurements were conducted using a combination of linear polarizers and quarter-wave plates. A fixed vertical analyzer was placed in front of the spectrometer to maintain a constant polarization orientation relative to the grating. Two quarter-wave plates, one in the incident beam path and one in the scattered beam path, were used to generate right- ($\sigma_R$) and left- ($\sigma_L$) circularly polarized light.

Single-crystal x-ray diffraction (XRD) patterns were collected at room temperature using a Bruker D8-Discover X-ray diffractometer with Cu K$\alpha_1$ radiation ($\lambda$ = 1.54 Å). Magnetic susceptibility and magnetization measurements were performed using a superconducting quantum interference device vibrating sample magnetometer (SQUID VSM, Quantum Design). Magnetoresistance and Hall effect measurements were performed at ambient pressure using the electrical transport option of the Quantum Design Physical Properties Measurement System (PPMS) with a conventional five-point contact configuration. The contacts were made with 10 μm gold wires and gold paint. The electrical current was applied in the *ab*-plane, while the magnetic field was applied along the crystallographic *c*-axis.

## III. RESULTS AND DISCUSSION

**A. Crystal structure, X-ray diffraction and electrical resistivity characterization**

Both pristine $CoNb_3S_6$ and $FeNb_3S_6$ crystallize in a non-centrosymmetric hexagonal structure (space group of *P*6$_3$22) [10,15]. Figure 1(a) shows the schematic representation of a crystal structure, in which the magnetic Co/Fe ions are intercalated between the $NbS_2$ layers.

Figure 1(b) illustrates the single-crystal x-ray diffraction patterns of $Co_{1-x}Fe_xNb_3S_6$ ($x$ = 0.1 and 0.3) measured at room temperature with the incident beam perpendicular to the (00l) planes. With increasing Fe concentration, a systematic shift of the (00l) reflections toward lower $2\theta$ angles is observed, indicating a slight elongation of the $c$-axis lattice parameter (11.94 Å for $x$ = 0.1 and 11.99 Å for $x$ = 0.3), driven by the larger atomic radius of Fe relative to Co, consistent with previous studies [27].

We further examine the electrical resistivity $\rho_{xx}$ of $Co_{1-x}Fe_xNb_3S_6$ across a temperature range of 2-300 K. As shown in Fig. 1(c), both compositions ($x$ = 0.1 and 0.3) display characteristic metallic behavior with a sudden drop below ~ 20.5 K for $x$ = 0.1 and 32.0 K for $x$ = 0.3, signaling the onset of long-rang magnetic ordering. To precisely identify the magnetic transition temperatures, we take the first derivative of the resistivity, $d\rho_{xx}/dT$ as a function of temperature, as shown in Fig. 1 (d). The sharp peaks in the $d\rho_{xx}/dT$ curves correspond to the Néel temperatures: $T_N$ ~ 20.5 K for $x$ = 0.1 and shifting to a higher $T_N$ ~ 32.0 K for $x$ = 0.3 (see the marked vertical lines). This shift indicates that Fe doping effectively tunes the magnetic exchange interactions, stabilizing the antiferromagnetic phase at higher temperatures. Additionally, the peak in the $d\rho_{xx}/dT$ is broader for $x$ = 0.3 than that of $x$ = 0.1. This broadening likely originates from increased structural or chemical disorder introduced by higher dopant concentrations, which can lead to inhomogeneous magnetic transitions and modified electron-electron scattering regimes.

**B. Lattice vibration and dynamics**

We next turn to Raman scattering measurements to probe phonon excitations. Factor group analysis for the $P6_322$ symmetry ($Z$=6 formulas per unit cell) predicts a total of 60 phonon modes at the Γ point with the following irreducible representation $\Gamma = 4A_1 + 6A_2 + 5B_1 + 5B_2 + 10E_1 + 10E_2$, see Table I for more details. Among them, $A_1$, $E_1$, and $E_2$ are Raman-active, while $E_1$ is normally forbidden in backscattering measurements. Figure 2(a) presents the unpolarized Raman spectra for $x = 0.1$ and $0.3$ collected at room temperature using 532 nm (2.33 eV) excitation. We observe five strong phonon excitations (labeled S1-S5) and one weak excitation (labeled P1) for both components. Interestingly as the Fe content increases from $x = 0.1$ to $0.3$, the 180 cm$^{-1}$ (S1) phonon exhibits a blue shift by $\sim 1.2$ cm$^{-1}$, while the other phonon modes remain largely unaffected (see the insets for S1 and S5 phonon excitations). This behavior indicates that the S1 phonon mode is selectively sensitive to the Fe doping, possibly related to the atomic mass difference between Fe and Co.

To identify the phonon symmetries and their polarization characteristics, we performed circularly polarized Raman measurements at room temperature. Figure 2(b) shows the circularly polarized Raman spectra for $x = 0.1$, along with unpolarized Raman spectra. The spectra were collected in co-circular ($\sigma_R\sigma_R$ or $\sigma_L\sigma_L$) and cross-circular ($\sigma_R\sigma_L$ or $\sigma_L\sigma_R$) polarized configurations. The S1, S2, S3, and S5 phonon excitations appear exclusively in co-circular polarized configuration, indicating $A_1$-type symmetry. In contrast, the P1 and S4 phonon excitations are observed under the cross-circular polarized configuration, suggesting $E_2$-type symmetry. These similar polarization characteristics were observed for the $x = 0.3$ sample as well (data not shown).

To validate these symmetry assignments, we evaluate the selection rules based on the Raman tensors for the $P6_322$ symmetry group. The polarization vectors of right- and left-handed circularly polarized light are given by $\sigma_R = [1 \ -i \ 0]/\sqrt{2}$ and $\sigma_L = [1 \ i \ 0]/\sqrt{2}$, respectively. Within the semiclassical approximation, the Raman scattering intensity for first-order phonon modes is expressed as $I_{int} = |\hat{e}_s^t . R . \hat{e}_i|^2$, where $R$ is the Raman tensor, and $\hat{e}_i$ and $\hat{e}_s$ are the incident and the scattered polarization vectors, respectively [28,29]. Using the Raman tensors listed in Table I, the scattering intensities for different phonon symmetries are obtained as follows: (i) For the $A_1$ phonon mode, the intensity in the co-circular polarization channels are given as $I_{A_1}(\sigma_R \sigma_R) = I_{A_1}(\sigma_L \sigma_L) \sim a^2$, but vanishes $I_{A_1}(\sigma_R \sigma_L) = I_{A_1}(\sigma_L \sigma_R) \sim 0$ in cross-circular channels. (ii) For the $E_1$ phonon mode, the scattering intensity is given as $I_{E_1}(\sigma_R \sigma_R) = I_{E_1}(\sigma_L \sigma_L) = I_{E_1}(\sigma_R \sigma_L) = I_{A_1}(\sigma_L \sigma_R) \sim 0$. (iii) For the $E_2$ phonon mode, the scattering intensities are given as $I_{E_2}(\sigma_R \sigma_R) = I_{E_2}(\sigma_L \sigma_L) \sim 0$ and $I_{E_2}(\sigma_R \sigma_L) = I_{E_2}(\sigma_L \sigma_R) \sim d^2$. From the above selection rules, it is evident that the $A_1$ and $E_2$ phonon excitations are allowed in the co-circular and cross-circular polarization configurations, respectively. In contrast, the $E_1$ modes remain forbidden in both circular polarization configurations.

We next inspect the temperature-dependent evolution of the phonon excitations. Figures 3 (a) and 3 (b) present 2D color contour maps of the Raman intensity versus Raman shift across a temperature window of 3.5 - 300 K for the $x = 0.1$ and $x = 0.3$ samples, respectively. All observed phonon modes exhibit a softening with increasing temperature. Intriguingly, the intensities of the S1-S4 phonon modes increase with increasing temperature, whereas the intensity of the S5 phonon mode decreases as the temperature rises. To quantify the phonon self-energy parameters, including the mode frequency, linewidth, and intensity, we extracted these quantities by fitting the Raman spectra using a sum of Lorentzian functions.

Figures 3 (c) and 3 (d) show the temperature-dependent frequencies (sphere, left axis) and intensities (square, right axis) of the few selected phonon excitations (S1, S4 and S5) for $x = 0.1$ and $x = 0.3$, respectively. We observe that the phonon excitations undergo a softening with increasing temperature particularly above $T_N$, which can be understood within the three-phonon anharmonicity decay model. In this framework, an optical phonon decays into two lower-energy phonons with equal energy but opposite momenta. Within the three-phonon anharmonicity model, the temperature-dependent frequency $\omega(T)$ is given by the expression $\omega(T) = \omega_0 + A(1+\frac{2}{e^x-1})$ [30], where $x = \hbar\omega_0/2k_B T$ and $\omega_0$ is the frequency of the phonon at 0 K. The coefficients A is a fitting parameter characterizing the strength of anharmonic phonon-phonon interactions. The solid red lines in Figs. 3 (c) and (d) represent the fits, which show good agreement with the experimental data. Notably, near and below $T_N$ (marked by the red vertical dashed line), we observe slight anomalies in the phonon frequencies prominent for the S1 and S5 phonon excitations, particularly for $x = 0.1$. These deviations from the anharmonic trend suggest an additional contribution probably arising from spin-phonon coupling [31–35]. In contrast, such anomalies are almost negligible for $x = 0.3$, indicating a reduced influence of magnetic ordering on the lattice dynamics or a smearing of spin–phonon coupling due to enhanced compositional or magnetic inhomogeneity.

Next, we turn to the temperature-dependent intensities of the phonon excitations, which exhibit pronounced mode- and composition-specific behaviors. As shown in Figs. 3 (c) and 3 (d), the intensities of the S1-S4 (intensities for S2 and S3 are not shown here) modes increase gradually with increasing temperature for both $x = 0.1$ and $x = 0.3$. In contrast, the intensity of the S5 phonon excitation shows a monotonic decrease with increasing temperature. The temperature evolution of Raman intensity is typically governed by a competition between the phonon population, the Raman scattering cross-section, and coupling to other degrees of freedom [28,29,33,36,37]. The gradual increase in the intensities of the S1-S4 phonon excitations is mainly attributed to the enhanced phonon population $n(\omega,T)$ governed by the Bose-Einstein statistics. Within the Bose-Einstein framework, the intensity of a first-order

Stokes Raman scattering can be expressed as $I \propto [n(\omega,T)+1]$. As the temperature increases, $n(\omega,T)$ increases, resulting in an enhancement of the Raman intensity. In contrast, the monotonic decrease in the intensity of the S5 mode suggests the involvement of additional damping channels, possibly arising from a combination of enhanced anharmonic phonon-phonon scattering and the coupling of this specific vibrational mode to magnetic fluctuations or spin-spin correlations [36].

For $x = 0.1$, subtle anomalies are seen in the intensity evolution of certain phonon excitations near and below $T_N$, see Fig. 3 (c). Such deviations from a smooth thermal trend suggest a modification of the Raman scattering cross section due to spin-phonon coupling. For $x = 0.3$, the intensity variations are comparatively smoother across $T_N$, see Fig. 3 (d), implying a weaker coupling between lattice vibrations and magnetic ordering for higher Fe doping.

## C. Magnetic susceptibility and magnetization

We turn to the magnetic properties of the Fe doped compounds. Figure 4 (a) shows the temperature-dependent magnetic susceptibility for $x = 0.1$, measured along the *ab*- plane ($\chi_{ab}$) and the *c*-axis ($\chi_c$). The measurements were conducted under an applied external field of $\mu_0 H$ = 0.1 T with both zero-field-cooled (ZFC) and field-cooled (FC) conditions. The ZFC and FC curves of $\chi_{ab}$ and $\chi_c$ exhibit nearly identical behavior across the entire temperature range.

Upon cooling, $\chi_c$ increases steeply with a subsequent drop at $T_N \sim 23.7$ K, as indicated by the red dashed line in Fig. 4 (a). This value is roughly 3.0 K higher than the $T_N$ derived from our resistivity measurements. Notably, the $T_N$ value obtained from both methods is lower than that reported for the parent compound, $CoNb_3S_6$ [25]. Unlike the parent compound $CoNb_3S_6$ [18], the $x = 0.1$ Fe-doped sample does not show a sharp increase in the FC $\chi_c$ near $T_N$. Nevertheless, $\chi_c$ remains consistently higher than $\chi_{ab}$ over the entire temperature range, consistent with previous studies on the parent material [18,38]. Meanwhile, $\chi_{ab}$ increases gradually with decreasing temperature and becomes nearly temperature-independent below $T_N$.

Figure 4 (b) shows the field dependence of $\chi_c$ for $x = 0.1$ under fields ranging from 0.1 T to 3 T. While the ZFC and FC curves follow similar trends at all field strengths, the overall magnitude of $\chi_c$ decreases as the magnetic field increases. A slight bifurcation between ZFC and FC curves is observed below ~15 K. Taken together, the observed anisotropic magnetic susceptibility is not typical for conventional easy $c$-axis antiferromagnets, suggesting the presence of complex, hidden magnetic correlations that extend beyond simple collinear antiferromagnetic ordering.

Figure 4 (c) illustrates the temperature dependence of the magnetic susceptibilities for the $x = 0.3$ composition measured under an applied field of $\mu_0 H = 0.1$ T. Similar to the $x = 0.1$ sample, the ZFC and FC curves for the in-plane magnetic susceptibilities are nearly identical. This is contrasted by the $\chi_c$ data, which shows a significant divergence occurs between the ZFC and FC data below $T_N$. Figure 4(d) details the field-dependent susceptibility along the $c$-axis for $x = 0.3$. Interestingly, below $T_N$, an increase in the applied magnetic field (from 0.1 T to 3 T) results in an increase in the magnitude of ZFC susceptibility, while the FC susceptibility decreases. We note that a ZFC-FC bifurcation below $T_N$ in an antiferromagnet is not inherent to a perfectly ordered antiferromagnet, and its presence may signal disorder, domain-wall pinning, spin canting, local moment inhomogeneity, or uniaxial anisotropy.

Next, we delve into the magnetization as a function of magnetic field and temperature. Figure 5 (a) and 5(b) show the $M-H$ curves for $x = 0.1$ and $x = 0.3$, respectively, measured at selected temperatures with the magnetic field applied along the $c$- axis. For $x = 0.1$, a noticeable hysteresis loop is observed at temperatures below $T_N$. Two distinct characteristic field-induced transitions, denoted as $H_{c1}$ and $H_{c2}$ are evident from the hysteresis loop, respectively. These transitions are assigned to the domain-switching field ($H_{c1}$, blue dashed line) and the metamagnetic transition ($H_{c2}$; red dashed line), following the previous reports on $Co_{1/3}TaS_2$ [11] and Fe-doped $Co_{1/3}NbS_2$ [27]. While $H_{c2}$ remains nearly temperature independent, $H_{c1}$ shifts rapidly toward lower fields with increasing temperature, merging with $H_{c2}$ around 18 - 20 K and approaching zero field near 24 K. At temperatures above 24 K, the hysteresis loop disappears and the magnetization varies linearly with the applied magnetic field ($M-H$ curves at some temperatures are not shown here). Remarkably, the hysteresis is more pronounced at 10 K than at 2 K. Below 10 K, the loop progressively weakens. This observation points to competing exchange interactions and possible magnetic phases within the antiferromagnetically ordered state, as inferred from the anomalous magnetic susceptibility unexpected for a simple antiferromagnet.

In contrast, the $M-H$ curves for $x = 0.3$ exhibit only a single, very weak hysteresis loop within a narrow field window of approximately –6 to +6 T. With increasing temperature, this field window progressively narrows and collapses completely around 20 K. Above 20 K, the magnetization varies linearly with the applied magnetic field. Notably, for $x = 0.3$, the hysteresis loop disappears well below $T_N$, whereas for $x = 0.1$, the hysteresis persists up to $T_N$.

Furthermore, upon careful examination, we find the separation trends in the $M$-$H$ curves for both $x = 0.1$ and 0.3. The separation between $M$-$H$ curves decreases for negative fields but increases for positive fields, see Fig. 5 (a) and 5 (b). Additionally, the magnitude of the hysteresis loop and the value of $M$ at higher fields are smaller at 2 K than␣10 K. This non-monotonic temperature dependence of the hysteresis suggests the presence of competing magnetic energy scales around 10 K.

**D. Magnetoresistance and Hall effect measurements**

To explore the interplay between charge transport and magnetic ordering, we investigated the magnetic-field dependence of magnetoresistance (MR) for $x = 0.1$ and 0.3. MR is defined as $\text{MR} = 100 \times [\rho(H) - \rho(0)] / \rho(0)$, where $\rho(0)$ and $\rho(H)$ represent the resistivity at zero field and under an applied magnetic field, respectively. Measurements were performed with the magnetic field applied along the $c$-axis and the electrical current applied within the $ab$-plane.

Figure 6 (a) presents the MR as a function of magnetic field at several representative temperatures. At temperatures above $T_N$, the MR curves are symmetric with respect to field reversal. The observed negative MR is typical for a paramagnetic state, where the application of a magnetic field suppresses spin-disorder scattering, thereby reducing the electrical resistivity. At low temperatures below $T_N$, a clear butterfly-shape of MR emerges, featuring distinct hysteresis between the forward ($-\mu_0 H$ to $\rightarrow \mu_0 H$; black) and reverse ($+\mu_0 H \rightarrow -\mu_0 H$; red) field sweeps. This behavior signifies strong coupling between charge carriers and magnetic ordering, likely driven by spin-dependent scattering associated with field-induced rearrangement of magnetic domains. The hysteresis is most prominent near 10 K. Upon further cooling to the lowest measured temperature of 2 K, the MR hysteresis is nearly suppressed, approaching the level observed at 20 K. However, while the 20 K profile is "M-shaped," the 2 K curve evolves into a distinct "W-shaped" structure. The positive MR at 2 K is due to electronic scattering caused by magnetic domain reorganization and Lorentz force effects. We stress that the non-monotonic MR behavior mirrors that of the $M$-$H$ curves, which show the strongest magnetic hysteresis at the same temperature. In contrast, for $x = 0.3$, no obvious hysteresis is observed at any temperature and the MR curves remain symmetric upon field reversal even well below $T_N$, see Fig. 6 (b). Upon cooling through $T_N$, the MR changes its sign from negative to positive.

To elucidate further the temperature evolution of the MR, we plot the MR at fixed magnetic fields of $\mu_0 H = \pm 9\,\text{T}$ as a function of temperature for $x = 0.1$ and 0.3 in Figs. 6 (c) and 6 (d), respectively. For $x = 0.1$, the MR exhibits a non-monotonic temperature dependence, showing a large negative MR around 10 K, consistent with the competing magnetic fluctuations and enhanced spin-scattering effects. For $x = 0.3$, the MR initially increases upon warming up to 10 K, and then decreases monotonically with further increase in temperature, eventually crossing into negative values at 50 K. Collectively, these results demonstrate that increasing the Fe concentration suppresses MR hysteresis and fundamentally changes the temperature-dependent charge transport; nevertheless, the hidden temperature scale $T^* \sim 10$ K continues to influence the magneto-transport properties even at higher doping levels.

Figure 7 (a) and 7 (b) present the magnetic-field dependence of the Hall resistivity ($\rho_{xy}$) measured at various selected temperatures above and below $T_N$ for $x = 0.1$ and $x = 0.3$, respectively. The magnetic field was applied along the $c$-axis and the electrical current flows within the $ab$-plane.

In magnetic materials, the Hall resistivity is typically described by the conventional relation $\rho_{xy} = \rho_{xy}^N + \rho_{xy}^{AHE} + \rho_{xy}^{THE} = R_H B + \rho_{xy}^{AHE} + \rho_{xy}^{THE}$, where the first term $\rho_{xy}^N = R_H B$ is the normal Hall effect contribution ($R_H$ is the ordinary Hall coefficient and $B = \mu_0 H$), the second term $\rho_{xy}^{AHE}$ is the anomalous Hall effect (AHE) contribution to the Hall resistivity and the third term $\rho_{xy}^{THE}$ is the topological Hall effect (THE), which arises from nontrivial spin textures. The AHE can be further expressed as $\rho_{xy}^{AHE} = R_s \mu_0 M$, where $R_s$ is the anomalous Hall coefficient and $M$ is the magnetization.

For *x* = 0.1, $\rho_{xy}$ shows a weak deviation from linearity on the applied magnetic field at temperatures below 24 K (except at 2K) and a small discernible hysteresis between the forward and reverse field sweeps. These features point to the presence of a weak AHE/THE contribution. Consistent with the magnetization data, this anomalous component progressively diminishes with increasing temperature and becomes negligible above 24 K. In contrast, for *x* = 0.3, $\rho_{xy}$ exhibits a linear dependence on the applied magnetic field and shows no discernible hysteresis between positive and negative field sweeps over the entire temperature range studied, see Fig. 7(b), suggesting that the charge transport in this composition is dominated primarily by the normal Hall effect.

To understand quantitatively the AHE/THE contributions, we estimated these components after the subtraction of the normal Hall contribution from the measured Hall resistivity data. We note that for the materials with very small net magnetization, the anomalous Hall contribution is expected to be weak. This is indeed the case for the present material, where the magnetization remains very small even at zero field. In our analysis, the ordinary Hall coefficient $R_H$ was extracted by performing a linear fit to the down-sweep Hall resistivity in the high-field region between 3 T and 9 T. The slope of this linear fit corresponds to the normal Hall coefficient. The normal Hall coefficient of the *x* = 0.1 sample is larger than that of the *x* = 0.3 sample, and for both samples, $R_H$ decreases monotonically with increasing temperature, see Fig. 8 (a).

After subtracting the linear ordinary Hall contribution from the total $\rho_{xy}$, the resulting topological Hall resistivity ($\rho_{xy}^{THE}$) is summarized in Figs 8 (b) for *x* = 0.1 and Fig 8 (c) for *x* = 0.3. For *x* = 0.1, a pronounced and hysteretic $\rho_{xy}^{THE}$ is observed at low temperatures except at 2 K, indicating a significant topological contribution to the Hall response. The magnitude of $\rho_{xy}^{THE}$ decreases systematically with increasing temperature and becomes negligible above 24 K. At zero field, the topological Hall resistivity exhibits a finite value except at 2 K. As the temperature increases, both the zero-field topological Hall resistivity and the coercive field systematically decrease, and the topological Hall signal is no longer observable at 30 K, see Figs. 8 (b) and 8 (f). At 2 K, the THE is not clearly observed, which contrasts with previous reports on Fe-doped $CoNb_3S_6$ in Ref. [27] while being consistent with Ref. [22]. The absence of a discernible topological Hall signal at 2 K is likely because the coercive field becomes larger than the maximum applied field of 9 T, preventing full alignment of magnetic domains with the same fictitious field direction and thus suppressing the net topological Hall response.

In contrast, for $x = 0.3$, the THE is much weaker and is barely discernible when the data are plotted using the same vertical scale as in Fig. 8 (b). To better visualize this weak signal for $x = 0.3$, Figs. 8 (d) and 8 (e) present enlarged views of $\rho_{xy}^{THE}$. Figure 8 (d) shows that a very small but finite topological Hall signal of approximately 0.02 µΩ.cm in the temperature range between 5 K and 18 K. At higher temperatures [Fig. 8 (e), 22–35 K], the signal further diminishes and approaches the noise level, indicating the disappearance of the topological Hall contribution.

In contrast to the $x = 0.1$ sample, the $x = 0.3$ composition exhibits a substantially larger magnetization. Moreover, the field dependence of the extracted Hall signal closely follows that of the corresponding *M-H* curves. This similarity suggests that the observed signal is more likely induced by the magnetization-related contribution, rather than originating from a genuine THE.

In the current study, increasing the Fe concentration likely introduces structural or magnetic disorder that modifies the magnetic texture and electronic landscape. This modification effectively reduces the Berry curvature or diminishes the specific non-collinear/non noncoplanar spin textures, leading to the suppression of anomalous transport observed in the $x = 0.3$ sample. Nonetheless, a remnant of complex non-collinear magnetic textures with the temperature scale of $T^*\sim 10$ K survives event at $x = 0.3$, leaving a discernible imprint on both the magneto-transport and magnetic behaviors. As such, localized non-collinear correlations or fluctuating magnetic clusters remain active around 10 K, influencing the scattering mechanisms and magnetic susceptibility of the system.

## IV. CONCLUSIONS

In summary, we systematically investigated the effects of Fe substitution on the phonon dynamics, magnetism, and charge transport properties of the intercalated vdW antiferromagnet $CoNb_3S_6$. Raman scattering measurements identified weak spin-phonon coupling near the Néel temperature $T_N$ for $x = 0.1$. Magnetic and magneto-transport characterizations confirm that the system maintains antiferromagnetic order characterized by strong $c$-axis anisotropy. However, Fe doping drives a significant reconstruction of the magnetic ground state. The $x = 0.1$ compound exhibits complex metamagnetic transitions, butterfly-shaped MR, and a weak topological Hall effect. In contrast, for $x = 0.3$, these anomalous transport phenomena are largely suppressed, although a remnant temperature scale of $T^*\sim 10$ K persists in the magnetic and transport behaviors. These results provide critical microscopic insights into the delicate interplay between lattice dynamics, magnetism, and charge transport in intercalated transition-metal dichalcogenides. Our findings underscore the high sensitivity of anomalous magneto-transport phenomena to subtle modifications in magnetic texture, even when only short-range or weak noncollinear magnetic correlations persist.

**Acknowledgments:**

This work was supported by the National Research Foundation (NRF) of Korea (Grant Nos. 2020R1A5A1016518 (RS-2020-NR049536), and RS-2023-00209121). R.S. acknowledges support from NSTC of Taiwan under No. 114-2124-M-001-009, 113-2112-M-001-045-MY3, and Academia Sinica Project No. AS-iMATE-115-11. R.K.U. would like to acknowledge the IITR for the Faculty Initiation Grant (No. FIG-101068). K. H. K. acknowledges the National Research Foundation (RS-2024-00338707) and the Ministry of Education through the core center program (2021R1A6C101B418).

**Data availability**

The data that support the findings of this article are not publicly available. The data are available from the authors upon reasonable request.

**Competing interests**

The authors declare no competing interests.

**Table I:** Wyckoff positions of the different atoms in the conventional unit cell and and irreducible representations of phonon modes, for CoNb$_3$S$_6$ with space group $P6_322$ #182 and point group $D_6$, at $\Gamma$ point of the Brilloun zone.

| Atom | Wyckoff site | Phonon mode decomposition at $\Gamma$ point |
|---|---|---|
| Co(2) | 2c | $A_1 + B_1 + E_1 + E_2$ |
| Nb(2) | 2a | $A_1 + B_2 + E_1 + E_2$ |
| Nb(4) | 4f | $A_1 + A_2 + B_1 + B_2 + 2E_1 + 2E_2$ |
| S (12) | 12i | $3A_1 + 3A_2 + 3B_1 + 3B_2 + 6E_1 + 6E_2$ |

**Total phonon modes at $\Gamma$ point**: $\Gamma_{Total} = 4A_1 + 6A_2 + 5B_1 + 5B_2 + 10E_1 + 10E_2$

**Raman tensor**:
$$R_{A_1} = \begin{pmatrix} a & 0 & 0 \\ 0 & a & 0 \\ 0 & 0 & b \end{pmatrix} \quad R_{E_1} = \begin{pmatrix} 0 & 0 & 0 \\ 0 & 0 & c \\ 0 & c & 0 \end{pmatrix}, \begin{pmatrix} 0 & 0 & -c \\ 0 & 0 & 0 \\ -c & 0 & 0 \end{pmatrix}$$

$$R_{E_2} = \begin{pmatrix} d & 0 & 0 \\ 0 & -d & 0 \\ 0 & 0 & 0 \end{pmatrix}, \begin{pmatrix} 0 & -d & -c \\ -d & 0 & 0 \\ -c & 0 & 0 \end{pmatrix}$$

**Figure:**

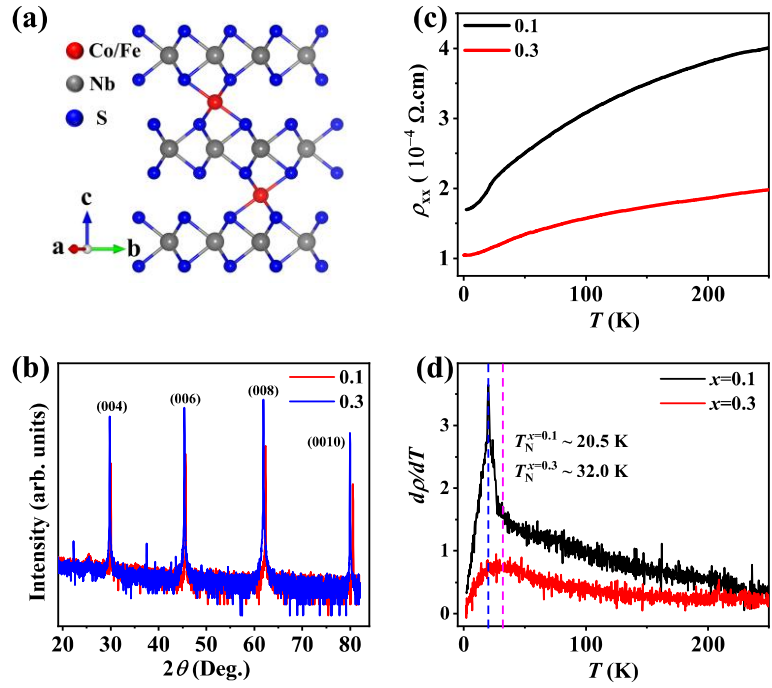

**Figure 1: (a)** Crystal and magnetic structure of end compound (Co/Fe)Nb$_3$S$_6$. **(b)** XRD patterns for the $x$ = 0.1 and 0.3. **(c)** Temperature-dependent longitudinal resistivity ($\rho_{xx}$). **(d)** Temperature-dependent first-derivative of the $\rho_{xx}$ for both $x$ = 0.1 and 0.3. The blue and pink dashed lines mark the $T_N$ for $x$= 0.1 and 0.3, respectively.

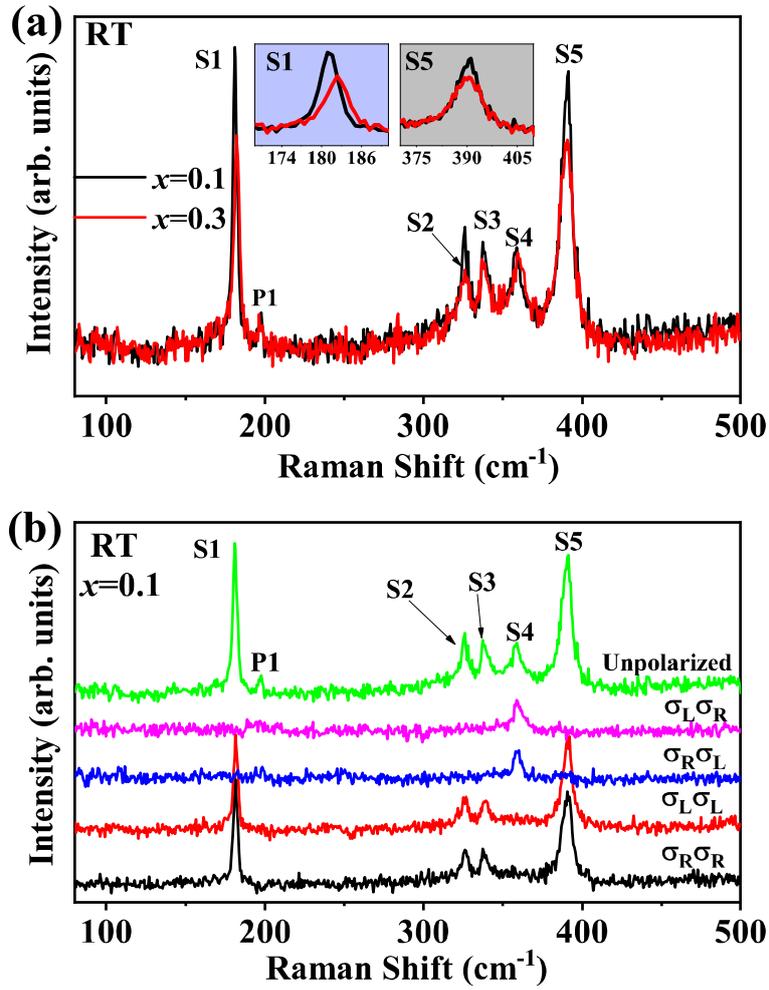

**Figure 2: (a)** Unpolarized Raman spectra for *x* = 0.1 (black) and 0.3 (red), and **(b)** Circularly polarized Raman spectra for *x* = 0.1, collected at room temperature using 532 nm (2.33 eV) excitation. The observed phonon excitations are labeled as S1-S5 and P1.

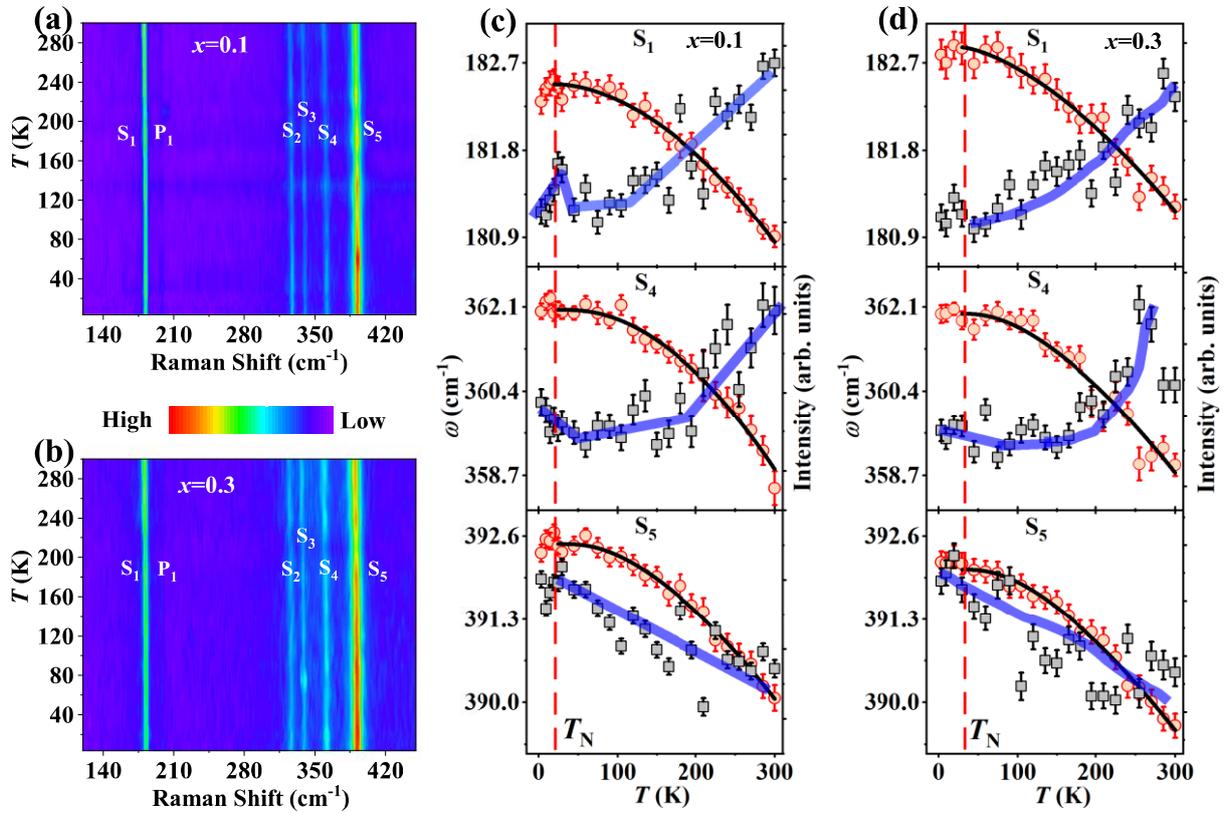

**Figure 3: (a)** and **(b)** Unpolarized 2D color contour maps of Raman intensity versus Raman shift as function of temperature for $x = 0.1$ and $x = 0.3$ respectively. **(c)** and **(d)** Temperature dependent frequency (sphere) and intensity (square) of the S1, S4 and S5 phonon excitations for $x = 0.1$ and $x=0.3$, respectively. Black solid lines are the fits using the Lorentzian function as described in the text. Semitransparent blue lines are guides to the eye. The red vertical dashed lines in **(c-d)** mark the $T_N$.

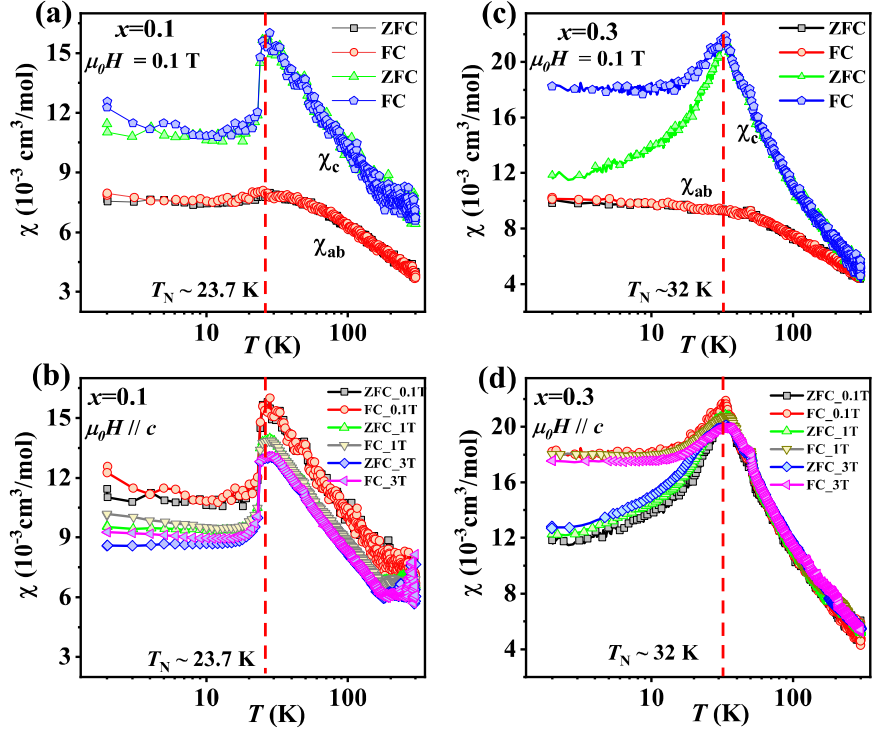

**Figure 4: (a)** and **(b)** Temperature-dependent magnetic susceptibility ($\chi$) for $x$= 0.1 and 0.3 along the ab plane ($\chi_{ab}$) and $c$-axis ($\chi_c$) measured under an external field of $\mu_0 H = 0.1$T and collected under ZFC and FC conditions, respectively. **(c)** and **(d)** $\chi - T$ for $x$ = 0.1 and 0.3 along the $c$-axis measured under an external field of $\mu_0 H = 0.1$T, 1T and 3T and collected under ZFC and FC conditions, respectively. The red vertical dashed lines in **(a-d)** marks the $T_N$.

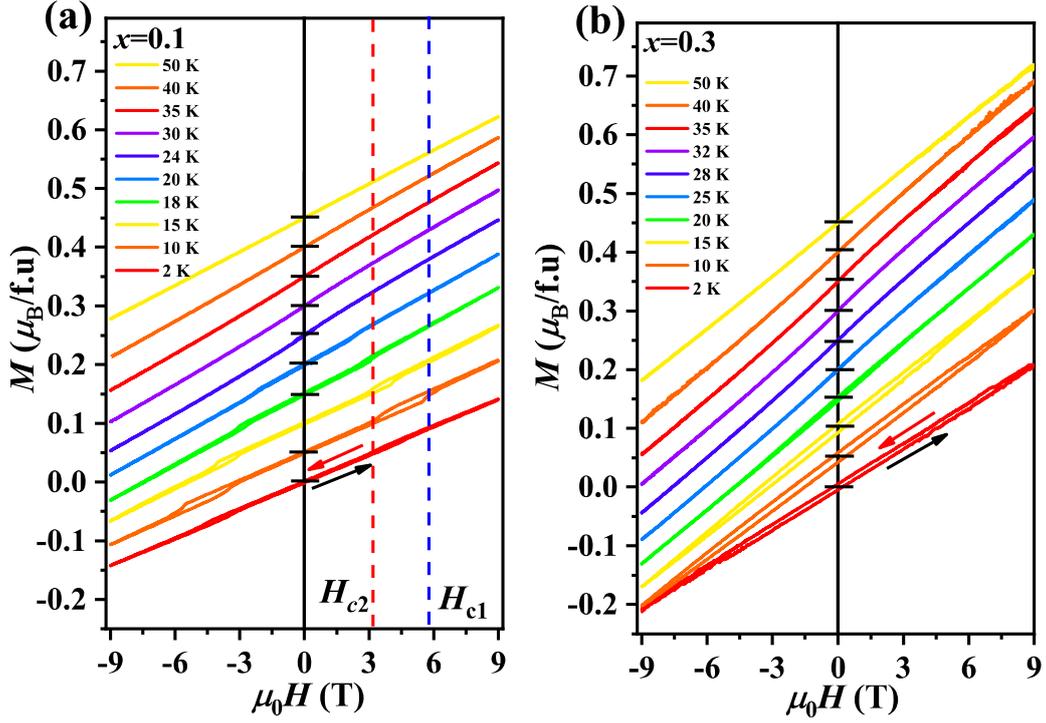

**Figure 5: (a)** and **(b)** Magnetization (*M*) versus magnetic field (*H*) at different temperatures for *x* = 0.1 and 0.3 respectively, with magnetic field applied parallel to the *c*-axis. The $M-H$ are vertically shifted for clarity. The dashed blue and red vertical lines in **(a)** mark the domain-switching field ($H_{c1}$) and the metamagnetic transition ($H_{c2}$), respectively. The black and red arrows in **(a)** and **(b)** indicate the magnetic-field directions from $-\mu_0 H$ to $+\mu_0 H$ and $+\mu_0 H$ to $-\mu_0 H$, respectively.

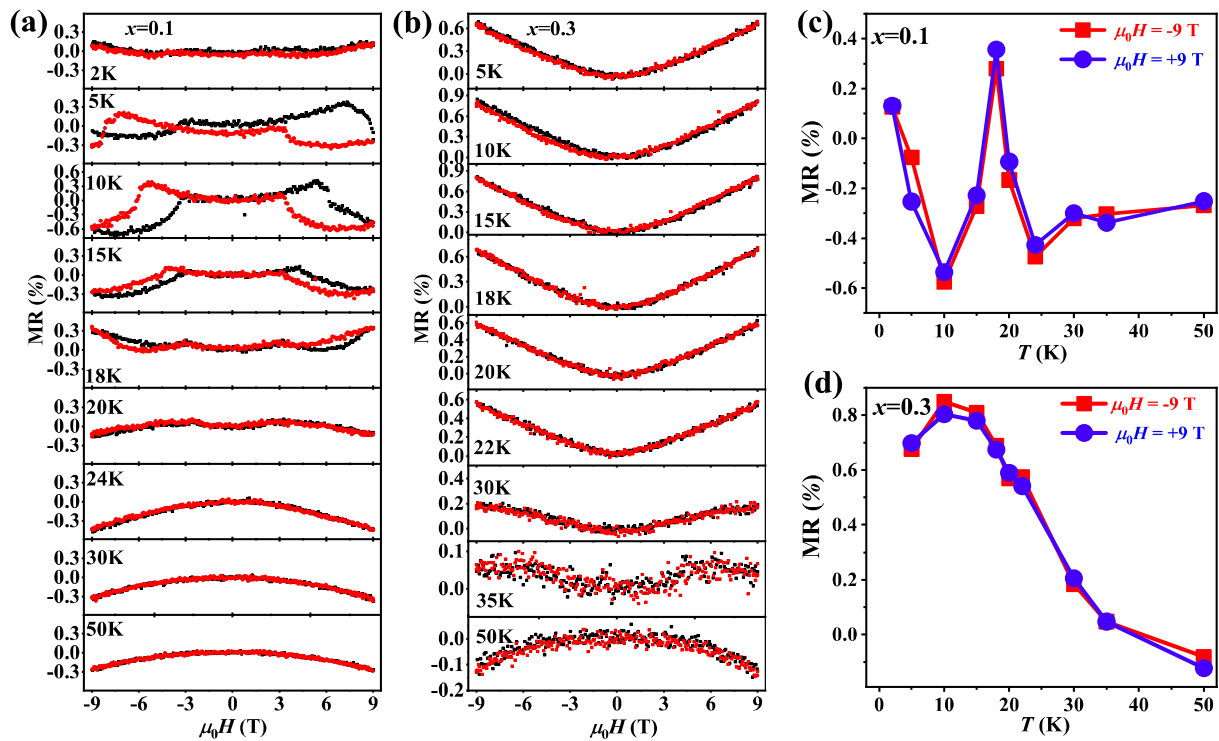

**Figure 6: (a)** and **(b)** Magnetic field dependence of magnetoresistance (MR) collected at different temperatures for *x* = 0.1 and 0.3 respectively, with the magnetic field applied along the *c*-axis and the electrical current was applied within the *ab*-plane. Black and red curve marks the magnetic-field directions from $-\mu_0 H$ to $+\mu_0 H$ and $+\mu_0 H$ to $-\mu_0 H$, respectively. **(c)** and **d)** Temperature dependence of the MR at magnetic field $\mu_0 H = \pm 9 \text{T}$ for *x* = 0.1 and 0.3, respectively.

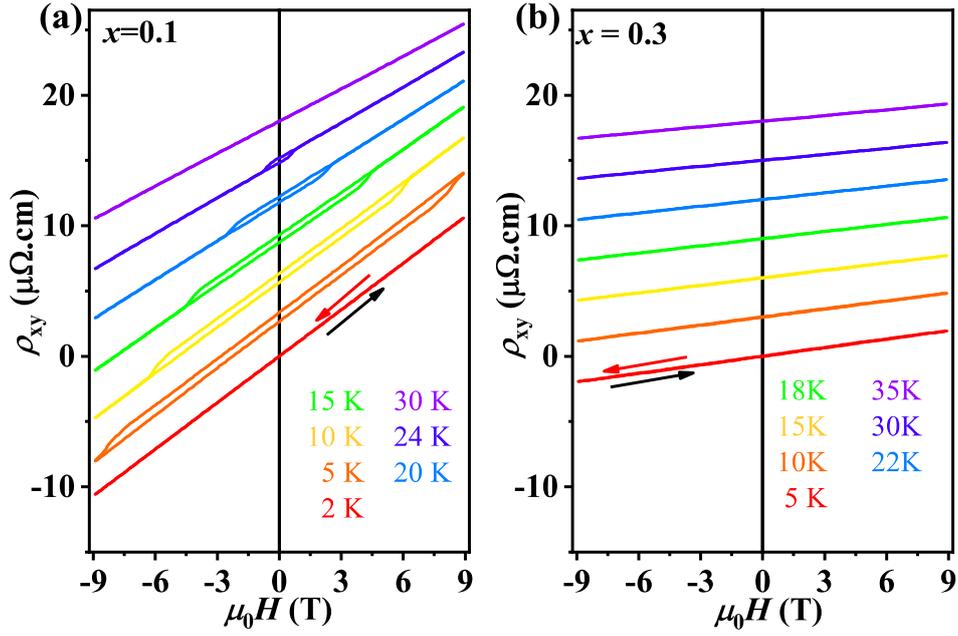

**Figure 7: (a)** and **(b)** Magnetic field dependence of Hall resistivity ($\rho_{xy}$) collected at different temperatures for $x$ = 0.1 and 0.3 respectively, with the magnetic field applied along the *c*-axis and the electrical current was applied within the *ab*-plane. Black and red curve marks the magnetic-field directions from $-\mu_0 H$ to $+\mu_0 H$ and $+\mu_0 H$ to $-\mu_0 H$, respectively.

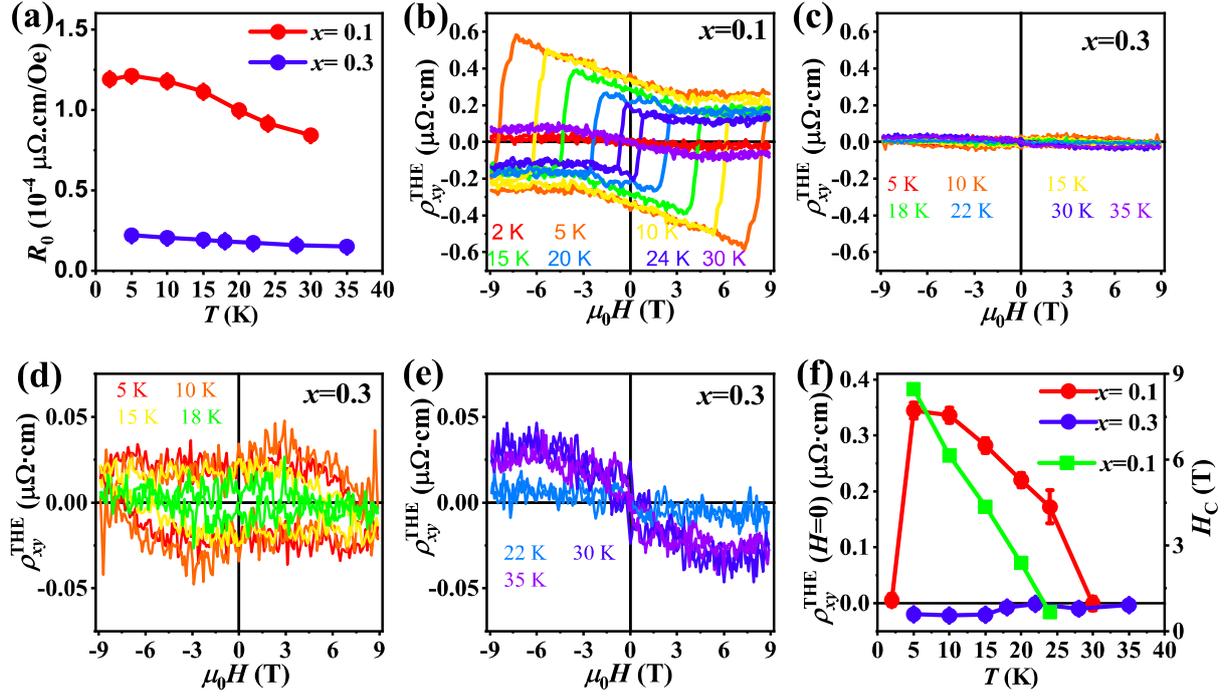

**Figure 8: (a)** Temperature dependence of the normal Hall coefficient ($R_H$) for $x = 0.1$ (red sphere), and $x = 0.3$ (blue sphere). **(b)** and **(c)** Topological Hall resistivity ($\rho_{xy}^{THE}$) for $x = 0.1$ and $x = 0.3$, respectively. **(d)** and **(e)** Amplified views of topological Hall resistivity for $x = 0.3$: **(d)** at 5 K, 10 K, 15 K, and 18 K; and **(e)** at 22 K 30 K, and 15 K. **(f)** Temperature dependence of the zero-field $\rho_{xy}^{THE}$ for $x = 0.1$ (red sphere), and $x = 0.3$ (blue sphere), and coercive field for $x = 0.1$ (green square).